\documentclass[aps,prb,twocolumn]{revtex4-2}
\usepackage{epsfig,amsopn,siunitx}
\usepackage{graphicx}
\usepackage{array}
\usepackage{physics}
\usepackage{color}
\usepackage{hyperref}
\hypersetup{
    colorlinks=true,
    linkcolor=magenta,
    citecolor=magenta,
    urlcolor=magenta
}
\usepackage{amsmath,amssymb}
\usepackage{enumerate}
\newcommand\bea{\begin{eqnarray}}
\newcommand\eea{\end{eqnarray}}
\newcommand\beq{\begin{equation}}
\newcommand\eeq{\end{equation}}

\newcommand{\noi}{\noindent}

\newcommand{\mcI}{\mathcal{I}}
\newcommand{\mcT}{\mathcal{T}}
\def\nn{\nonumber}
\def\f{\frac}

\def\si{\sigma}

\begin{document}

\title{Electrically controlled spin-splitting and asymmetric tunnel magnetoresistance in anti-altermagnets}
\author{Abhiram Soori} 
\email{abhirams@uohyd.ac.in}
\affiliation{School of Physics, University of Hyderabad, Prof. C. R. Rao Road, Gachibowli, Hyderabad 500046, India}

\begin{abstract}
Anti-altermagnets (AAMs) are a recently identified class of layered magnetic materials where opposite spin-splitting in adjacent layers creates a globally spin-degenerate band structure, rendering conventional spectroscopic detection highly difficult. In this paper, we theoretically demonstrate an all-electrical method to manipulate and probe this hidden magnetic order using a dual-gated transport junction. By applying a perpendicular displacement field, we break the spatial inversion symmetry of the lattice, explicitly lifting the global spin degeneracy. We attach ferromagnetic (FM) leads on either side of the AAM. Using quantum transport calculations, we show that this gate-induced spin-splitting manifests as a strongly asymmetric tunnel magnetoresistance (TMR) as a function of the lead magnetization. We identify specific crystallographic orientations where the TMR  retains its symmetry despite the fully split bands, a direct consequence of exact momentum-space compensation. We further reveal that Rashba spin-orbit coupling guarantees robust, highly directional transport asymmetries even in the absence of an explicit  chemical potential mismatch between the layers. Our findings establish a clear, electrically tunable framework for exploiting AAMs in next-generation spintronic architectures.
\end{abstract}

\maketitle

\noi{\it Introduction .--}
Altermagnets represent an unconventional magnetic phase characterized by a spin-split band structure despite possessing zero macroscopic net magnetization~\cite{smejkal22b,sukhachov2024,soori2026prm}. This unique configuration makes them highly attractive for spintronic applications~\cite{Fu2025}, as they eliminate parasitic stray magnetic fields while still supporting spin-polarized currents alongside standard charge transport~\cite{das2023}. Furthermore, tunnel magnetoresistance (TMR) in altermagnetic junctions exhibits strong crystallographic orientation-dependent characteristics~\cite{Sun25,Ezawa2026,ghadi2026}, offering additional degrees of freedom for device design.

More recently, anti-altermagnets (AAMs) have emerged as a distinctive layered variant of these materials~\cite{aam2025prl,aam2026prl,lange2026}. AAMs are bilayer altermagnetic systems wherein the spin-splitting in one layer is perfectly compensated by an opposite splitting in the adjacent layer. This results in a globally spin-degenerate band structure, rendering conventional spectroscopic identification of this magnetic order inherently difficult.  The combined operation of layer inversion ($\mcI$) and time reversal ($\mcT$) preserves the symmetry of the system, even though the individual operations $\mcI$ and $\mcT$ break it. While applying an in-plane magnetic field has been proposed as a mechanism to separate these oppositely polarized bands~\cite{sun2026aam}, purely electrical control remains highly desirable. Several promising material candidates for AAMs have already been identified, including {Cs$_{1-\delta}$V$_2$Te$_2$O}~\cite{yang2026aam} and ${\mathrm{Sr}}_{n+1}{\mathrm{Cr}}_{n}{\mathrm{O}}_{3n+1}$~\cite{aam2026prl}. 

\begin{figure}
\includegraphics[width=\columnwidth]{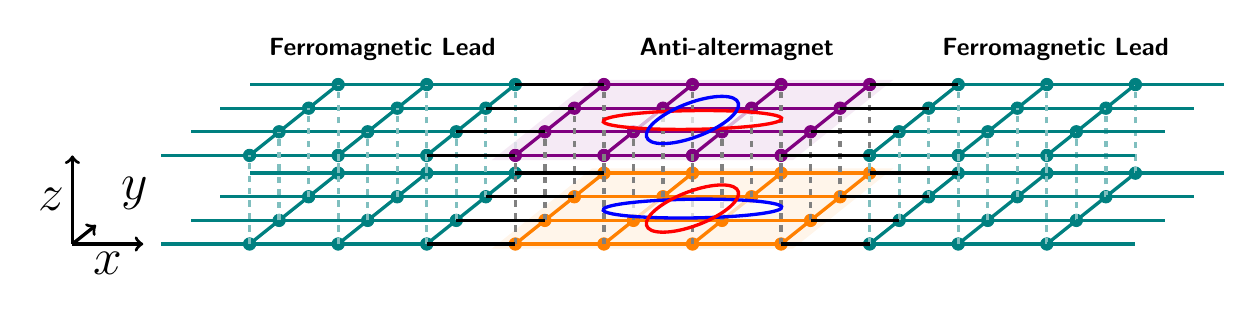}
\caption{ Schematic diagram of the transport setup,  by attaching ferromagnetic leads to the central AAM region. A dual-gate configuration is employed to apply independent top and bottom gate voltages to the AAM. Consequently, the otherwise globally spin-degenerate band structure is electrically split, manifesting as differently sized Fermi surfaces for opposite spins in the adjacent layers. }\label{fig:schem}
\end{figure}

In layered two-dimensional systems such as graphene and transition metal dichalcogenides, dual-gating is a well-established experimental technique known to unlock rich functional phase diagrams~\cite{Zhang2009,pablo2010,abdullah2023,hu2025}. In this paper, we investigate the electronic and transport properties of dual-gated AAMs. We show that independent top and bottom gating generates a perpendicular displacement field that breaks $\mcI\mcT$, successfully splitting the otherwise degenerate AAM bands. We attach ferromagnetic leads to both sides of the dual-gated AAM [as shown in Fig.~\ref{fig:schem}] to probe its transport signatures. Using Landauer formalism applied to a lattice model, we demonstrate that the TMR manifests as an asymmetric function of magnetization whenever the bands are electrically split. Remarkably, we also identify special crystallographic orientations of the AAM where, even in the fully spin-split regime, the TMR retains perfect symmetry about zero magnetization.

{\noindent \it Calculations .--} Because the system maintains translational invariance along the $y$-direction, the transverse momentum $k_y$ is a conserved quantum number. Consequently, the total tight-binding Hamiltonian can be decoupled into independent momentum sectors and expressed as $H = \int dk_y H_{k_y} / 2\pi$, where
\begin{widetext}
\bea
H_{k_y} &=& -\sum_{n_x=-\infty}^{0}\big[ t (c^{\dagger}_{n_x-1}c_{n_x} +{\rm h.c.})-c^{\dagger}_{n_x}\{t_{\perp} \sigma_0\tau_x    -(2t \cos k_y +\mu_l)\si_0\tau_0-b\si_{\beta,\gamma}\tau_0\} c_{n_x}\big]  \nn \\ 
&& -t' (c^{\dagger}_{1}c_{0} +{\rm h.c.}) -\sum_{n_x=1}^{L-1}\big[ c^{\dagger}_{n_x+1}\{ t\sigma_0\tau_0 +t_J( \cos{2\phi}  -i\sin{2\phi}\sin{k_y})\si_z\tau_z -\f{i\alpha}{2}\sigma_y\tau_0 \}c_{n_x}+ {\rm h.c.} \big] \nn \\ 
&& -\sum_{n_x=1}^{L}c^{\dagger}_{n_x}\big[ (2t\cos{k_y}+\mu_a)\sigma_0\tau_0 -\alpha\sin{k_y}\sigma_x\tau_0 -2t_J\cos{2\phi}\cos{k_y}\sigma_z\tau_z+\Delta\mu\sigma_0\tau_z -t_{\perp}\sigma_0\tau_x \big] c_{n_x} \nn \\ 
&& -t'(c^{\dagger}_{L+1}c_{L}+{\rm h.c.}) -\sum_{n_x=L+1}^{\infty}\big[ t (c^{\dagger}_{n_x+1}c_{n_x} +{\rm h.c.}) - c^{\dagger}_{n_x}\{t_{\perp}\sigma_0\tau_x  -(2t \cos k_y +\mu_l)\si_0\tau_0-b\si_{\beta,\gamma}\tau_0\} c_{n_x}\big] .  
\eea
\end{widetext}
Here, $\sigma_j$ and $\tau_j$ denote the Pauli matrices acting on the spin and layer subspaces, respectively, and $\si_{\beta,\gamma}=\cos\beta\si_z+\sin\beta(\cos\gamma\si_x+\sin\gamma\si_y)$. While the Neel vector of AAM is taken to be along $\hat z$ direction, the spin polarisation of FM leads points in a direction that makes spherical polar angles $(\beta,\gamma)$. The parameters $t$ and $t_{\perp}$ represent the intralayer and interlayer kinetic hopping amplitudes. The local chemical potentials in the ferromagnetic leads and the central region are given by $\mu_l$ and $\mu_a$, respectively, while $b$ characterizes the Zeeman energy that dictates the spin polarization in the leads. The interfacial coupling between the leads and the central region is parameterized by the hopping strength $t'$. Furthermore, $t_J$ defines the strength of the altermagnetic exchange, $\phi$ specifies the crystallographic orientation angle relative to the $x$-axis, and $\Delta\mu$ represents the interlayer chemical potential mismatch due to dual gating.

The dispersion relation in the FM leads is given by
\begin{equation}
E_{\tau,\sigma} = -2t(\cos k_x+\cos k_y) +\tau t_{\perp} -\sigma b,
\end{equation}
where $\tau=\pm$ and $\sigma=\pm$ correspond to the eigenvalues of the layer Pauli matrix $\tau_x$ and the spin Pauli matrix $\sigma_{\beta,\gamma}$, respectively. In the central anti-altermagnet region, the dispersion relation takes the form
\begin{eqnarray}
E &=& -2t(\cos k_x+\cos k_y)-\mu_a\pm\sqrt{t_{\perp}^2+d_z^2}, \quad \text{where} \nonumber \\
d_z &=& -2\si t_J[\cos(2\phi)(\cos k_x-\cos k_y) \nn \\ && +\sin(2\phi)(\sin k_x\sin k_y)] -\Delta\mu,
\end{eqnarray}
in the limit of $\alpha=0$. 
The propagating wavenumber in the $y$-direction is restricted to the range $(-k_{\tau,\sigma,0}, k_{\tau,\sigma,0})$, where the boundary is defined by $k_{\tau,\sigma,0}=\cos^{-1}\left[-(E+\mu_l-\tau t_{\perp}+\sigma b+2t)/2t\right]$ for $\tau=\pm$ and $\sigma=\pm$, provided that the resulting $k_{\tau,\sigma,0}$ is strictly real. Consequently, the scattering eigenfunction for an incident wave at energy $E$ in a specified channel $(\tau,\sigma)$ with transverse wavenumber $k_y$ is given by $\psi_{\tau,\si} (n_x)e^{ik_yn_y}$, where 
\begin{widetext}
\bea
\psi_{\tau,\si} (n_x) &=& \begin{cases} e^{ik_{x\tau\si}n_x} \ket{\tau,\si}+ \sum\limits_{\tau',\si'} r_{\tau'\si';\tau\si} e^{-ik_{x\tau'\si'}n_x} \ket{\tau',\si'}{\rm ~for~} n_x \le 0 \\ 
\sum\limits_{\tau',\si'} t_{\tau'\si';\tau\si} e^{ik_{x\tau'\si'}n_x}\ket{\tau',\si'} {\rm ~for~} n_x \ge L+1
\end{cases}
\eea
\end{widetext}

Here, $k_{x\tau'\si'}$ are calculated from the dispersion relation for the band $\tau',\si'$ in the FM lead keeping $k_y$ same throughout. The differential conductivity is then given by 
\bea 
G(E, b) &=& G_0\sum\limits_{\tau\si\tau'\si'}^{\prime} \int_{-k_{0\tau\si}}^{k_{0\tau\si}} \f{d{k_y}}{2\pi} \f{\sin{[{\rm Re}(k_{x\tau'\si'})]}}{\sin{k_{x\tau\si}}} \f{|t_{\tau'\si';\tau\si}|^2}{ g(k_y)}, \nn \\ 
&& {\rm where ~~}G_0=e^2/h, {\rm ~~and} \nn \\ g(k_y) &=& \sqrt{1-\Big[\f{(E+\mu_l+\si b-\tau t_{\perp})}{2t}+\cos{k_y}\Big]^2}
\eea 
The prime on the summation explicitly restricts the transport evaluation  to physically propagating channels, ensuring that only those bands yielding a purely real wavevector in the leads $k_{0\tau\sigma}$ are included in the calculation. The magnetoresistance is then given by 
\bea
MR(b) &=& \f{G(E,0)-G(E,b)}{G(E,b)}
\eea
In rest of the paper we fix $E=0$ (Fermi energy) and calculate MR as a function of $b$. 

{\noindent \it Parameter choice.--} We adopt the following set of parameters to obtain our numerical results: $t_J=0.6t$, $t_{\perp}=0.5t$, $\mu_l=\mu_a=-0.2t$, and $t'=0.8t$, guided by choices made in previous studies on anti-altermagnets~\cite{sun2026aam,yang2026aam}. Furthermore, we choose $L=10$ throughout our transport calculations.

\begin{figure}
\includegraphics[width=0.45\columnwidth]{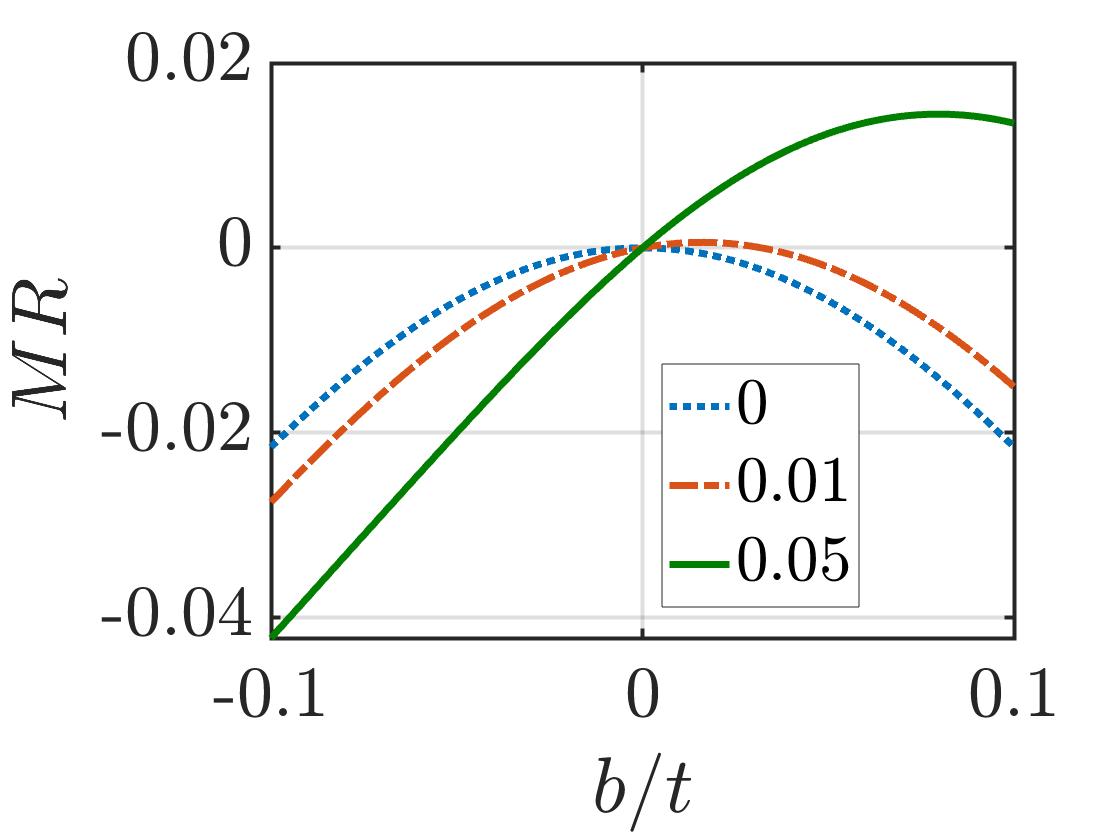}
\includegraphics[width=0.45\columnwidth]{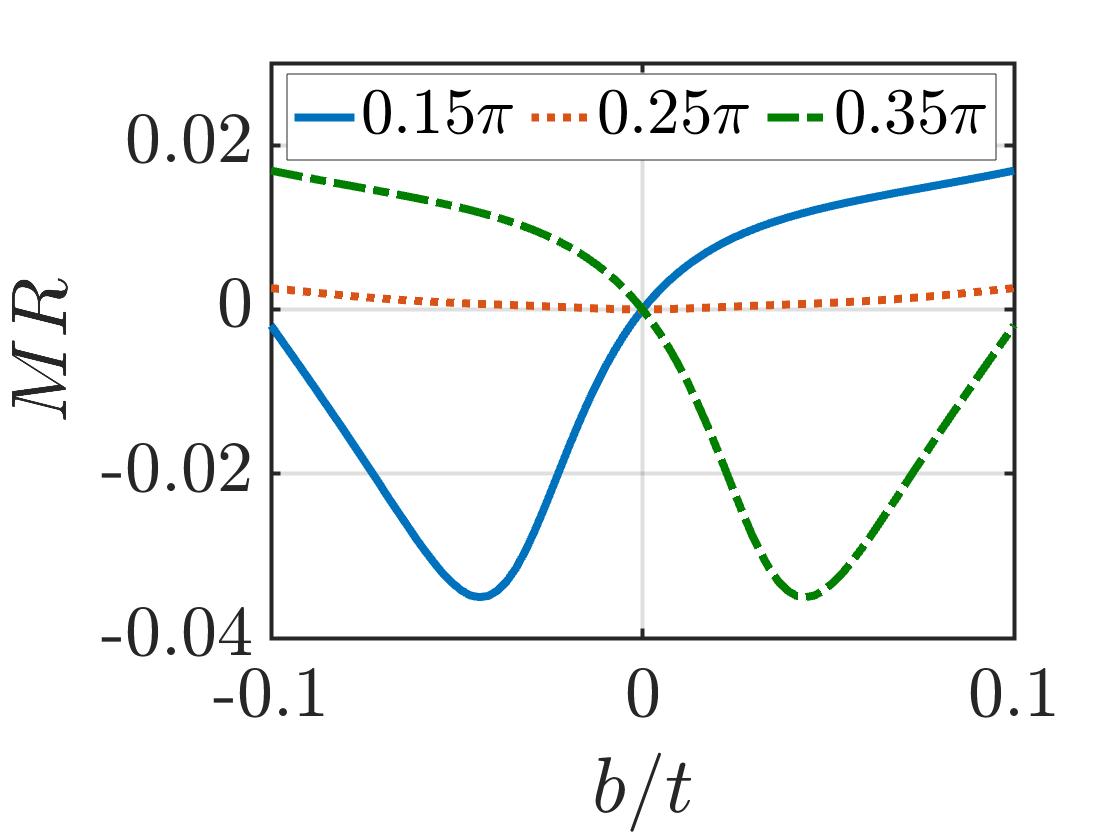}
\includegraphics[width=0.45\columnwidth]{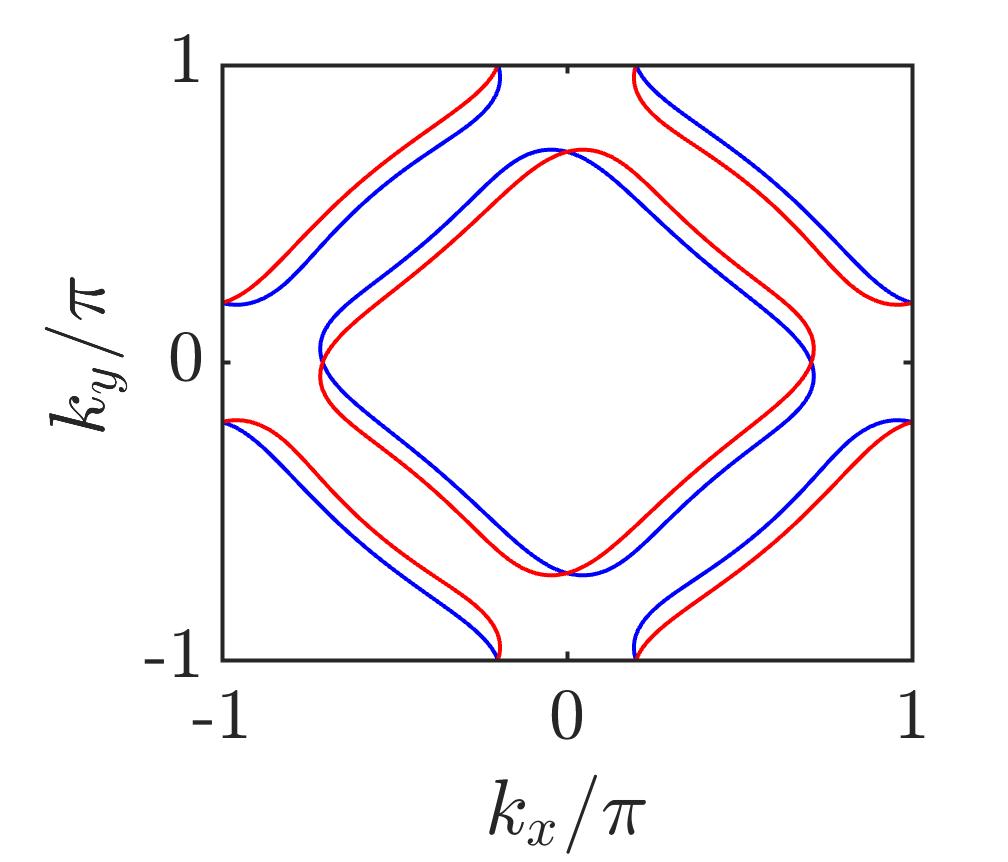}
\includegraphics[width=0.45\columnwidth]{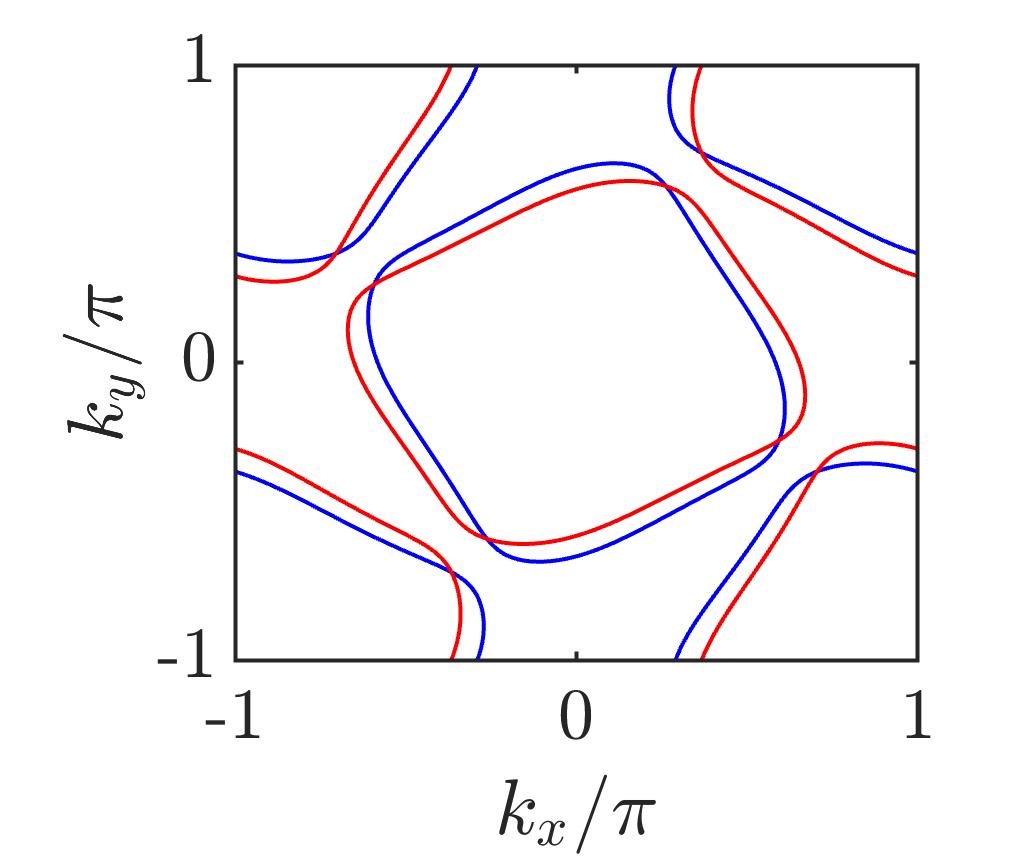}
\caption{(a,b) MR as a function of the Zeeman field strength, which parameterizes the magnetization of the ferromagnetic leads. (c,d) Momentum-space Fermi surface contours of the central AAM region. In (a), the MR curves are evaluated for varying values of the chemical potential mismatch $\Delta\mu$ at a fixed crystallographic orientation $\phi=0$. In (b), the curves illustrate the MR dependence on varying orientations $\phi$ at a fixed  $\Delta\mu=0.1t$. Panels (c) and (d) display the Fermi contours for $\phi=0.25\pi$ and $\phi=0.15\pi$, respectively, for $\Delta\mu=0.3t$. This enhanced value of $\Delta\mu$, compared to panels (a) and (b), is selected to visually resolve the band spin-splitting induced by the broken spatial inversion symmetry. Other parameters: $t_J=0.6t$, $t_{\perp}=0.5t$, $\mu=-0.2t$, $t'=0.8t$ and $L=10$. }
\label{fig:res1}
\end{figure}

{\noindent \it  Asymmetric MR .--} In Figure~\ref{fig:res1}(a,b), we plot MR versus the Zeeman field, which quantifies the magnetization of the FM leads. As illustrated in Figure~\ref{fig:res1}(a), applying a nonzero dual-gate voltage--parameterized by the induced chemical potential mismatch $\Delta\mu$--breaks the spatial inversion symmetry of the lattice. Consequently, the symmetric transport signature $MR(-b)=MR(b)$ ceases. This asymmetric behavior holds true for any crystallographic orientation $\phi$ that is not an odd multiple of $\pi/4$. Physically, this phenomenon arises because a finite $\Delta\mu$ disrupts the perfect degeneracy of the system; the Fermi surfaces for the two spins no longer compensate one another. As a result, the AAM effectively behaves as two coupled altermagnetic monolayers with their spin-splitting profiles mutually rotated by $\pi/2$. Because the mismatched sizes of the Fermi contours prevent exact cancellation of spin polarization, one altermagnetic layer dominates the overall spin-dependent quantum transport resulting in asymmetric $MR$ versus $b$ plot. 
In Figure~\ref{fig:res1}(b), we plot MR versus $b$ for different $\phi$  and observe that it remains invariant under $b \to -b$ at the specific crystallographic orientation $\phi=\pi/4$, even in the presence of a finite chemical potential mismatch $\Delta\mu$. As $\phi$ deviates from this axis, the symmetric MR profile vanishes. Instead, it obeys a generalized  relation: $MR(b,\phi=\pi/4-\theta)=MR(-b,\phi=\pi/4+\theta)$. This behavior is fundamentally rooted in the AAM Hamiltonian, which satisfies the transformation property $H(t_J,k_y,\phi=\pi/4-\theta)=H(-t_J,-k_y,\phi=\pi/4+\theta)$. The physical origin of this transport signature is visually evident in Figure~\ref{fig:res1}(c); at $\phi=\pi/4$, every state on the up-spin Fermi contour is mapped to a corresponding down-spin state with identical $k_x$ but opposite $k_y$, ensuring a symmetric MR with respect to the Zeeman field. As illustrated in Figure~\ref{fig:res1}(d), this strict momentum-space compensation is absent when $\phi$ deviates from $\pi/4$, yielding an asymmetric MR versus $b$ profile. Nevertheless, the Hamiltonian's intrinsic symmetry  guarantees the broader relation $MR(b,\phi=\pi/4-\theta)=MR(-b,\phi=\pi/4+\theta)$ across the parameter space.

\begin{figure}
\includegraphics[width=0.48\columnwidth]{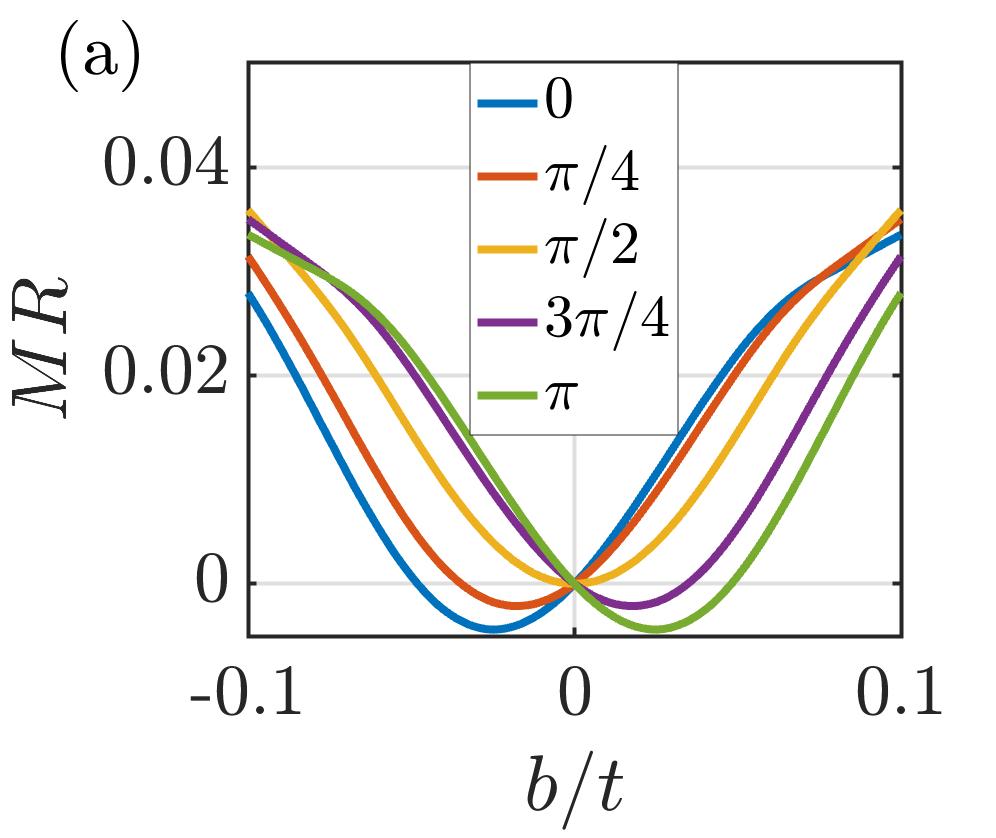}
\includegraphics[width=0.48\columnwidth]{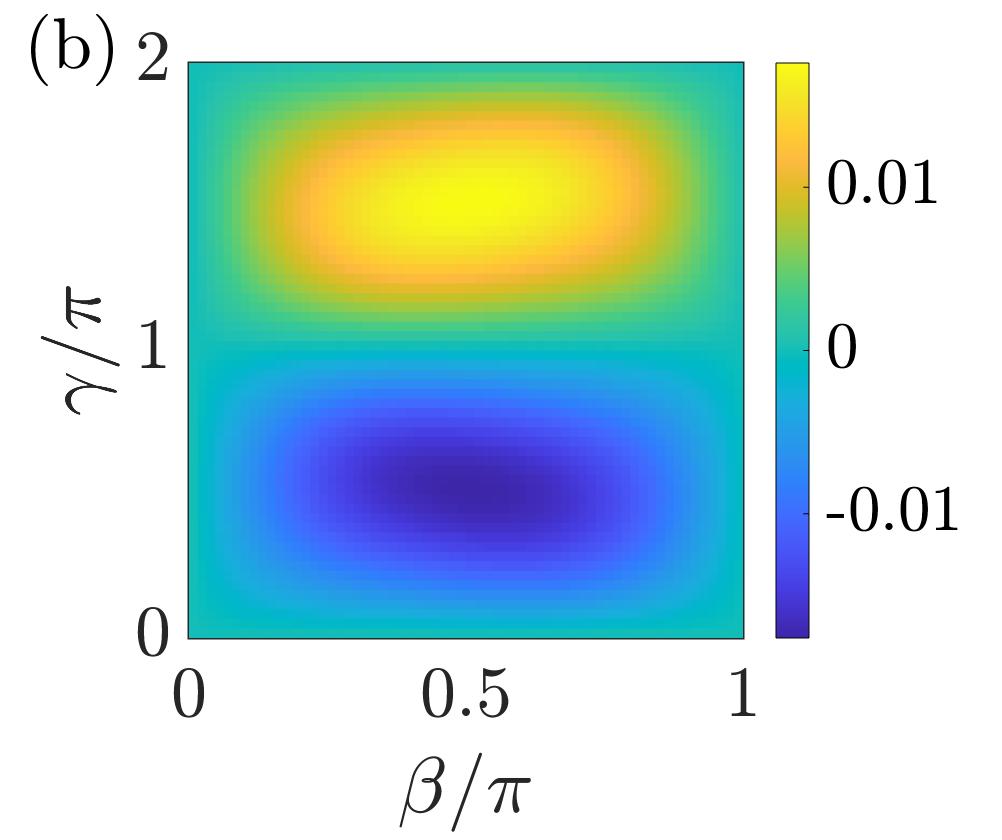}
\caption{(a) MR as a function of the Zeeman field $b$ for various choices of the angle $\beta$, defined between the N\'eel vector of the AAM and the spin polarization direction of the FM leads, evaluated at a finite chemical potential mismatch $\Delta\mu = 0.05t$. (b) Dependence of $\Delta MR/\Delta b$ on the spherical polar angles $(\beta,\gamma)$ relative to the N\'eel vector of the AAM in the absence of a chemical potential mismatch ($\Delta\mu = 0$). All other parameters are identical to those in Fig.~\ref{fig:res1}(a).}
\label{fig:res2}
\end{figure}
{\noindent \it Misaligned spin easy-axes .--} Next, we investigate the configuration where the magnetization of the ferromagnetic leads is not parallel to the N\'eel vector of the AAM. Figure~\ref{fig:res2}(a) presents the calculated MR as a function of the Zeeman field $b$ for various angles $\beta$, defined between the ferromagnetic spin polarization axis and the N\'eel vector of AAM. The MR profile exhibits a  symmetry about $b=0$ when $\beta=\pi/2$. More broadly, for any $\beta$ within the range $[0,\pi]$, the transport satisfies the generalized relation $MR(b,\beta)=MR(-b,\pi-\beta)$. Physically, this  symmetry arises because the MR is dictated by the relative alignment between the N\'eel vector of AAM and the spin polarization direction of FM; applying the transformation $b \to -b$ concurrently with $\beta \to \pi-\beta$ perfectly maps the parameter space $(b,\beta)$ back onto itself.

{\noindent \it  Rashba spin-orbit coupling .--}
In Figure~\ref{fig:res2}(b), we plot the differential asymmetry $\Delta MR/\Delta b \equiv [MR(b)-MR(-b)]/2b$, evaluated at $b=0.1t$, as a function of the spherical polar angles $(\beta,\gamma)$. This configuration is analyzed in the absence of a chemical potential mismatch ($\Delta\mu=0$) but in the presence of a finite Rashba spin-orbit coupling (SOC) strength $\alpha = 0.1t$. We observe that $\Delta MR/\Delta b$ exhibits a pronounced extrema near $(\beta,\gamma)=(\pi/2,\pi/2), (\pi/2,3\pi/2)$, corresponding to a state where the FM leads are spin-polarized  along the $y$-axis. This behavior originates directly from the Rashba SOC term in the AAM Hamiltonian, which explicitly couples the in-plane momentum $k_x$ to the Pauli spin matrix $\sigma_y$. Consequently, injecting $y$-polarized spins creates momentum-dependent scattering asymmetries across the Fermi contours, yielding a highly asymmetric MR. Crucially, if the altermagnetic exchange is disabled ($t_J=0$), the central region reduces to a standard time-reversal symmetric SOC system. 
Under these conditions,  time-reversal symmetry of SOC region guarantees that the transmission probabilities for opposite spin directions perfectly compensate one another across the Brillouin zone, restoring a perfectly symmetric MR profile and forcing $\Delta MR/\Delta b = 0$~\cite{Sahoo2023}.
It is therefore the intrinsic altermagnetic term that provides the essential time-reversal symmetry breaking, in addition to inversion symmetry breaking along $x$ by SOC,  ensuring a robust, nonzero $\Delta MR/\Delta b$ even in the complete absence of a chemical potential mismatch between the two layers.

{\noi \it Summary and Conclusion .--} We have proposed an all-electrical framework to  manipulate and probe the hidden magnetic order in AAMs utilizing a dual-gated FM-AAM-FM device architecture. By applying a perpendicular displacement field to break the spatial inversion symmetry of the lattice, we explicitly lift the global spin degeneracy via a tunable chemical potential mismatch. Our quantum transport calculations demonstrate that this symmetry breaking manifests as a strongly asymmetric TMR with respect to the Zeeman field $b$. Crucially, this transport signature is strictly governed by the underlying crystal symmetries; perfectly symmetric TMR is  preserved at specific crystallographic orientations (e.g., $\phi = \pi/4$) due to exact momentum-space compensation between the spin-split Fermi contours. Furthermore, we established that injecting spins  in the presence of Rashba SOC triggers pronounced, highly directional transport asymmetries. Because the intrinsic altermagnetic exchange provides the mandatory time-reversal symmetry breaking, these directional transport signatures remain robust even when the explicit interlayer chemical potential mismatch is absent. Ultimately, these unique, gate-tunable transport properties provide a distinct non-spectroscopic pathway for identifying anti-altermagnetic phases and lay the groundwork for exploiting them in purely electrical, spintronic applications free from stray-magnetic-fields.

\noi {\it Acknowledgments.-- } 
The author thanks  Amit Agarwal for  timely discussions. The author acknowledges the financial support received from the Anusandhan National Research Foundation (erstwhile Science and Engineering Research Board) under the Core Research Grant (No. CRG/2022/004311), and the University of Hyderabad. 

\bibliography{ref_aam}
\end{document}